\documentclass[fleqn,twoside]{article}
\usepackage{espcrc2}
\usepackage{axodraw}
\usepackage{color}

\readRCS
$Id: espcrc2.tex,v 1.2 2004/02/24 11:22:11 spepping Exp $
\usepackage{graphicx}
\usepackage[figuresright]{rotating}

\newcommand{\AmS}{{\protect\the\textfont2
  A\kern-.1667em\lower.5ex\hbox{M}\kern-.125emS}}

\def\beq{\begin{equation}}
\def\eeq{\end{equation}}
\def\beeq{\begin{eqnarray}}
\def\eeeq{\end{eqnarray}}
\def\nn{\nonumber}

\def\res{{\rm Res}_{ \{{\rm Im}\; q_0 < 0 \}}}
\def\td{{\widetilde \delta}}

\title{From multileg loops to trees (by-passing Feynman's Tree Theorem)
\thanks{Work supported by Ministerio de Ciencia e Innovaci\'on under
grants FPA2007-60323 and CPAN (CSD2007-00042), by the European Commission
under contracts FLAVIAnet (MRTN-CT-2006-035482), 
HEPTOOLS (MRTN-CT-2006-035505), and MCnet (MRTN-CT-2006-035606), 
by the INFN-MEC agreement and by BMBF.
}}

\author{
Germ\'an Rodrigo \address[IFIC]{
   Instituto de F\'{\i}sica Corpuscular,
   CSIC-Universitat de Val\`encia,\\
   Apartado de Correos 22085, E-46071 Valencia, 
   Spain}\thanks{E-mail: german.rodrigo@ific.uv.es},
Stefano Catani \address[FI]{
   INFN, Sezione di Firenze and Dipartimento di Fisica, 
   Universit\`a di Firenze,\\ I-50019 Sesto Fiorentino, Florence, 
   Italy}\thanks{E-mail: stefano.catani@fi.infn.it},
Tanju Gleisberg \address[SLAC]{
   Stanford Linear Accelerator Center, Stanford University,
   Stanford, CA 94309, 
   USA}\thanks{E-mail: tanju@slac.stanford.edu},
Frank Krauss \address[IPPP]{
   Institute for Particle Physics Phenomenology, 
   Durham University, Durham DH1 3LE, 
   UK}\thanks{E-mail: frank.krauss@durham.ac.uk}, and  
Jan-C. Winter \address[FNL]{
   Fermi National Accelerator Laboratory, Batavia, IL 60510, 
   USA}\thanks{E-mail: jwinter@fnal.gov}
}
       
\begin{document}

\begin{abstract}
We illustrate a duality relation between one-loop integrals and 
single-cut phase-space integrals. The duality relation
is realised by a modification of the customary $+i0$ prescription of the
Feynman propagators. The new prescription regularizing the propagators,
which we write in a Lorentz covariant form, compensates for the absence
of multiple-cut contributions that appear in the Feynman Tree Theorem.
The duality relation can be extended to generic one-loop quantities, such as
Green's functions, in
any relativistic, local and unitary field theories.
\end{abstract}

\maketitle

\setcounter{footnote}{0}

\section{INTRODUCTION}

The physics program of LHC requires the evaluation of 
multi-leg signal and background processes at 
next-to-leading order (NLO). In the recent years, important 
efforts have been devoted to the calculation of 
many $2\to 3$ processes and some $2\to 4$ processes
(see, e.g., \cite{Bern:2008ef}).

We have recently proposed a method \cite{Catani:2008xa,meth,tanju} to 
compute multi-leg one-loop cross sections in perturbative field 
theories. The method uses combined analytical and numerical techniques.
The starting point of the method is a duality relation 
between one-loop integrals and phase-space integrals. In this respect,
the duality relation has analogies with the Feynman's Tree Theorem 
(FTT) \cite{Feynman:1963ax}. 
The key difference with the FTT 
is that the duality relation involves only single 
cuts of the one-loop Feynman diagrams.
In this talk, we illustrate the duality relation, and 
discuss its correspondence, similarities, and differences
with the FTT.

\section{NOTATION}

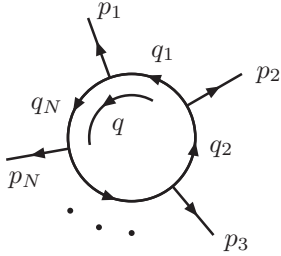
\begin{figure}[tb]
\begin{center}
\begin{picture}(100,110)(0,-10)
\SetScale{.8}
\SetWidth{1.2}
\BCirc(50,50){30}
\ArrowArc(50,50)(30,110,190)
\ArrowArc(50,50)(30,190,-50)
\ArrowArc(50,50)(30,-50,30)
\ArrowArc(50,50)(30,30,110)
\ArrowArc(50,50)(20,60,190)
\ArrowLine(39.74,78.19)(29.48,106.38)
\ArrowLine(75.98,65)(101.96,80)
\ArrowLine(69.28,27.01)(88.56,4.03)
\ArrowLine(20.45,44.79)(-9.09,39.58)
\Vertex(21.07,15.53){1.4}
\Vertex(34.60,7.71){1.4}
\Vertex(50,5){1.4}
\Text(35,44)[]{$q$}
\Text(32,88)[]{$p_1$}
\Text(52,72)[]{$q_1$}
\Text(92,64)[]{$p_2$}
\Text(74,36)[]{$q_2$}
\Text(8,52)[]{$q_N$}
\Text(0,24)[]{$p_N$}
\Text(80,0)[]{$p_3$}
\end{picture}
\vspace{-1.2cm}
\caption{\label{f1loop}
{Momentum configuration of the one-loop $N$-point scalar integral.}}
\end{center}
\end{figure}

We consider a generic one-loop integral $L^{(N)}$ 
with massless internal lines (Fig.~\ref{f1loop}):
\beq
\label{Ln}
L^{(N)}(p_1, p_2, \dots, p_N) = 
\int_q \;\;
\prod_{i=1}^{N} \, \frac{1}{q_i^2+i 0}~, 
\eeq
where $q^\mu$ is the loop momentum (which flows anti-clockwise), and we use 
the shorthand notation:
\beq 
\int_q \;\; \dots \equiv - i \, \int \frac{d^d q}{(2\pi)^d}\;\; \dots \;\;. 
\nonumber
\eeq 
The momenta of the external legs, which are taken as outgoing
and are clockwise ordered, are denoted 
by $p_1^\mu, p_2^\mu, \dots, p_N^\mu$,  
with $\sum_{i=1}^N p_i = 0$, and $p_{N+i} \equiv p_{i}$. 
The momenta of the internal lines are given by 
$q_i = q + \sum_{k=1}^i p_k$.

\section{THE FEYNMAN TREE THEOREM}
\label{sec:ft}

The FTT \cite{Feynman:1963ax} applies to any
local and unitary quantum field theory in Minkowsky space with an
arbitrary number $d$ of space-time dimensions. It relates perturbative
scattering amplitudes and Green's functions at the loop level with
analogous quantities at the tree level. 

Let us introduce the customary Feynman ($G$)
and advanced ($G_A$) propagators:
\beq
G(q) \equiv \frac{1}{q^2+i0}~,\qquad G_A(q) \equiv \frac{1}{q^2-i0\,q_0}~,
\eeq
where $q_0$ is the energy component of the $d$-dimensional momentum $q_\mu$.
They are related by
\beq
\label{gavsg}
G_A(q)=G(q)+\td(q)~,
\eeq
with $\td(q) \equiv 2 \pi  i \, \theta(q_0) \, \delta(q^2) 
= 2 \pi i \, \delta_+(q^2)$.
In the complex plane of $q_0$ both propagators have two poles. 
The pole of the Feynman propagator 
with positive (negative) energy is slightly displaced below (above) 
the real axis, while both poles of the advanced propagator are located 
above the real axis. Hence, by using the Cauchy residue theorem in the $q_0$ complex 
plane, with the integration contour closed at $\infty$ in the lower 
half-plane, the one-loop integral of advanced propagators vanishes: 
\beq
\label{lna}
\int_q \;\; \prod_{i=1}^{N} \,G_A(q_i)=0~.
\eeq 
Inserting Eq.~(\ref{gavsg}) in Eq.~(\ref{lna}), and collecting 
all the terms with an equal number of delta functions, we obtain 
\beq
\label{lnavsln} 
\!0 = \!\!\int_q \prod_{i=1}^{N} \left[
G(q_i)+\td(q_i) \right] 
=\!  L^{(N)} + \sum_{m=1}^N L_{\rm{m-cut}}^{(N)}~.
\eeq
The $m$-cut integrals $L_{\rm{m-cut}}^{(N)}$ are the 
contributions with precisely $m$ delta functions:
\beeq
\label{lmcut}
&& \!\!\!\!\!\!\!\!
L_{\rm{m-cut}}^{(N)}(p_1, p_2, \dots, p_N) = \int_q \bigg\{
\td(q_1) \cdots \td(q_m) \nn \\ && \times \,
G(q_{m+1}) \cdots G(q_{N}) + {\rm uneq. \; perms.} \bigg\}~,
\eeeq
where the sum in the curly bracket includes all the permutations of
$q_1,\dots,q_N$ that give unequal terms in the integrand. 
From Eq.~(\ref{lnavsln}), we thus derive the 
FTT:
\beq
\label{lftt}
L^{(N)} = - \sum_{m=1}^N 
L_{\rm{m-cut}}^{(N)}~.
\eeq
The FTT relates the one-loop integral $L^{(N)}$ to 
the multiple-cut\footnote{
	If the number of space-time dimensions is $d$, the right-hand side 
	of Eq.~(\ref{lftt}) receives contributions only from the terms with 
	$m \leq d$.}
integrals $L_{\rm{m-cut}}^{(N)}$. Each delta function $\td(q_i)$ in 
$L_{\rm{m-cut}}^{(N)}$ replaces the corresponding Feynman propagator 
in $L^{(N)}$ 
by `cutting' the internal line with momentum $q_i$. Here, `cutting' is synonymous to setting
the respective particle on shell. An $m$-particle cut decomposes the one-loop 
diagram in $m$ tree diagrams: in this sense, the FTT allows us to calculate 
loop diagrams from tree-level diagrams.

The extension of the FTT from one-loop integrals $L^{(N)}$ to one-loop
scattering amplitudes ${\cal A}^{({\rm 1-loop})}$ (or Green's functions) in
perturbative field theories is straightforward, provided the corresponding
field theory is unitary and local. The generalization of 
Eq.~(\ref{lftt}) to arbitrary scattering amplitudes is 
~\cite{Feynman:1963ax}:
\beq
\label{aftt}
{\cal A}^{({\rm 1-loop})} = - \sum_{m=1}^N 
{\cal A}^{({\rm 1-loop})}_{\rm{m-cut}}~,
\eeq
where ${\cal A}^{({\rm 1-loop})}_{\rm{m-cut}}$ is obtained 
by considering all possible replacements 
of $m$ Feynman propagators $G(q_i)$ of internal loop lines in 
${\cal A}^{({\rm 1-loop})}$ with $m$ on-shell propagators $\td(q_i)$.

\section{A DUALITY THEOREM}
\label{sec:dt}

In this Section we present and 
illustrate the duality relation between one-loop
integrals and single-cut phase-space integrals~\cite{Catani:2008xa}.

Applying the residue theorem directly  
to $L^{(N)}$, we obtain
\beeq
\label{ln4}
&& \!\!\!\!\!\!\!\!\!\!\!\!
L^{(N)}(p_1, p_2, \dots, p_N)  \nn \\ 
&& = - \,2 \pi i \; \int_{\bf q} 
 \;\;\sum \; \res
 \;\left[ \;\prod_{i=1}^{N} \,G(q_i) \right]~.  \nn \\
\eeeq 
The integral does not vanish (unlike the case of advanced propagators) 
since the Feynman propagators produces $N$ 
poles in the lower half-plane that contribute 
to the residues in Eq.~(\ref{ln4}).
The calculation of these residues is elementary, but it involves several
subtleties \cite{Catani:2008xa}. We get 
\beq
\label{resGi}
{\rm Res}_{\{i^{\rm th}{\rm pole}\}} \;\frac{1}{q_i^2+i0} 
= \int dq_0 \;\delta_+(q_i^2)~.
\eeq
This result shows that considering the residue of the Feynman propagator 
of the internal line with momentum $q_i$ is equivalent to cutting that 
line by including the 
corresponding on-shell propagator $\delta_+(q_i^2)$. 
The other propagators $G(q_j)$, with $j\neq i$, which are not 
singular at the value of the pole of $G(q_i)$, contribute as 
follows~\cite{Catani:2008xa}:
\beeq
\label{respre}
&& \!\!\!\!\!\!\!\!\!\!\!\!
\left. \prod_{j\neq i} \,\frac{1}{q_j^2 + i0} \,  
\right|_{q_i^2=-i0}
= \prod_{j\neq i} \; \frac{1}{q_j^2 - i0 \,\eta (q_j-q_i)}~, \nn \\ &&
\eeeq
where $\eta$ is a future-like vector, 
\beq
\label{etadef}
\eta_\mu = (\eta_0, {\bf \eta})~, \quad \eta_0 \geq 0~, 
\quad \eta^2 = \eta_\mu \eta^\mu \geq 0~,
\eeq
i.e.~a $d$-dimensional vector that can be either light-like $(\eta^2=0)$ or 
time-like $(\eta^2 > 0)$ with positive definite energy $\eta_0$.

We see from Eq.(\ref{respre}) that the calculation of 
the residue at the pole of the $i^{\rm th}$ 
internal line modifies the $i0$ prescription of the propagators  
of the other internal lines of the loop. This modified regularization is named
`dual' $i0$ prescription, and the corresponding propagators are named
`dual' propagators. The dual prescription arises from the 
fact that the original Feynman propagator $1/(q_j^2 +i0)$ 
is evaluated at the {\em complex} value of the loop momentum $q$, 
which is determined by the location of the pole at $q_i^2+i0 = 0$. 

The presence of $\eta$ is a consequence of the 
fact that the residue at each of the poles is not a Lorentz-invariant 
quantity. A given system of coordinates has to be specified to apply 
the residue theorem. Different choices of the future-like vector 
$\eta$ are equivalent to different choices of the coordinate system.
The Lorentz-invariance of the loop integral is recovered 
after summing over all the residues.

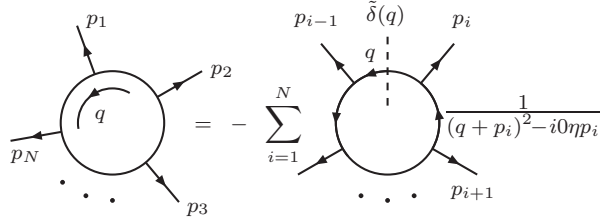
\begin{figure}[tb]
\begin{center}
\vspace{-.8cm}
\begin{picture}(350,110)(0,-10)
\footnotesize
\SetScale{.65}
\SetPFont{Times}{20}
\SetWidth{1.2}
\BCirc(50,50){30}
\ArrowArc(50,50)(20,60,190)
\ArrowLine(39.74,78.19)(29.48,106.38)
\ArrowLine(75.98,65)(101.96,80)
\ArrowLine(69.28,27.01)(88.56,4.03)
\ArrowLine(20.45,44.79)(-9.09,39.58)
\Vertex(21.07,15.53){1.4}
\Vertex(34.60,7.71){1.4}
\Vertex(50,5){1.4}
\Text(27,71)[]{$p_1$}
\Text(74,52)[]{$p_2$}
\Text(0,20)[]{$p_N$}
\Text(65,0)[]{$p_3$}
\Text(28,35)[]{$q$}
\Text(84,32)[]{$\displaystyle =~-~\sum_{i=1}^N$}
\SetOffset(104,0)
\BCirc(50,50){30}
\ArrowArc(50,50)(30,90,130)
\ArrowArc(50,50)(30,-30,50)
\ArrowArc(50,50)(30,130,220)
\ArrowLine(69.28,72.98)(88.57,95.96)
\ArrowLine(30.71,72.98)(11.43,95.96)
\ArrowLine(75.98,35)(101.96,20)
\ArrowLine(24.02,35)(-1.96,20)
\DashLine(50,60)(50,100){5}
\Text(6,71)[]{$p_{i-1}$}
\Text(61,71)[]{$p_{i}$}
\Text(65,6)[]{$p_{i+1}$}
\Text(26,58)[]{$q$}
\Text(32,74)[]{$\tilde{\delta}(q)$}
\Text(84,35)[]{$\frac{\displaystyle 1}
{\displaystyle (q+p_i)^2 \!\! - \! i0\eta p_i}$}
\Vertex(50,5){1.4}
\Vertex(65.39,7.71){1.4}
\Vertex(34.6,7.71){1.4}
\end{picture}
\vspace{-1cm}
\caption{\label{fig:dt}
{The duality relation for the one-loop $N$-point scalar integral:
graphical representation as a sum of $N$ basic dual integrals.}}
\end{center}
\end{figure}

The insertion of the results of Eqs.~(\ref{resGi})-(\ref{respre}) in
Eq.~(\ref{ln4}) gives us the duality relation
between one-loop integrals and single-cut phase-space integrals~\cite{Catani:2008xa}:
\beq
\label{ldt}
L^{(N)} = - \; {\widetilde L}^{(N)}~,
\eeq
where the explicit expression of the phase-space 
integral ${\widetilde L}^{(N)}$ is (cf. Fig.~\ref{fig:dt}) 
\beeq
\label{dcut}
&& \!\!\!\!\!\!\!\!\!
{\widetilde L}^{(N)}(p_1, p_2, \dots, p_N) \nn \\
&& = \int_q \;\; \sum_{i=1}^N
\; \td(q_i) \;
\prod_{\stackrel {j=1} {j\neq i}}^{N} 
\; \frac{1}{q_j^2 - i0 \,\eta (q_j-q_i)}~.
\eeeq
Each of the $N-1$ propagators in the integrand is regularized 
by the dual $i0$ prescription. 

Using the invariance of the integration measure under momentum 
translations in arbitrary $d$ dimensions, we can also rewrite 
Eq.~(\ref{dcut}) as a sum of $N$ 
basic phase-space integrals:
\beeq
\label{d1cutsum}
&& \!\!\!\!\!\!\!\!
{\widetilde L}^{(N)}(p_1, p_2, \dots, p_N) \nn \\
&=& \sum_{i=1}^N \;
I^{(N-1)}(p_i, p_{i,i+1}, \dots, p_{i,i+N-2})~,
\eeeq 
where $p_{i,j}=p_i + p_{i+1} + \ldots + p_j$.
The basic one-particle phase-space integrals with 
$n$ dual propagators are denoted 
by $I^{(n)}$, and are defined as follows:
\beeq
\label{idual}
&& \!\!\!\!\!\!\!\!\!
I^{(n)}(k_1, k_2, \dots, k_n) \nn \\ && =
\int_q \td(q) \;\prod_{j=1}^{n} \;
\frac{1}{2qk_j + k_j^2 - i0 \,\eta k_j} \,.
\eeeq

Summarizing our results, we find that:
\begin{itemize}
\item
The multiple-cut contributions $L_{\rm{m-cut}}^{(N)}$, with $m \geq 2$, 
of the FTT are completely absent from the duality relation, 
which involves single-cut contributions only.
\vspace*{-2mm} 
\item
The Feynman propagators in ${L}^{(N)}$ are replaced by dual 
propagators in ${\widetilde L}^{(N)}$. 
\vspace*{-2mm}
\item
The dual $i0$ prescription and the basic dual integrals $I^{(n)}$
depend on the auxiliary vector $\eta$. 
However, ${\widetilde L}^{(N)}$ does not depend on $\eta$,
provided it is fixed to be the same in all its contributing single-cut
terms (dual integrals).
\end{itemize}

The expression (\ref{d1cutsum}) of ${\widetilde L}^{(N)}$ as a sum 
of basic dual integrals is actually a single phase-space integral, 
whose integrand is the sum of the terms obtained by cutting each 
of the internal loop lines. 
The duality relation, therefore, directly expresses
the one-loop integral as the phase-space integral of a tree-level 
quantity.
In the case of the FTT, the relation between loop and tree-level quantities 
is more involved, 
since the multiple-cut contributions $L_{\rm{m-cut}}^{(N)}$ (with $m \geq 2$) 
contain integrals of expressions that correspond to the product of $m$
tree-level diagrams over the phase-space for different number of particles.

\section{RELATING THE FTT WITH THE DUALITY THEOREM}

The FTT and the duality theorem can be related in a direct way starting
from a basic identity between dual and Feynman 
propagators~\cite{Catani:2008xa}:
\beeq
\label{dovsfp}
&& \!\!\!\!\!\!\!\!\!\!\!\!
\td(q) \, \frac{1}{2qk + k^2 - i0 \,\eta k} \nn \\ &&
= \td(q) \;\Bigl[ G(q+k) + \theta(\eta k) \;\td(q+k) \Bigr]~. 
\eeeq
This identity applies to the dual propagators when they are inserted in a
single-cut integral.
The proof of the equivalence of the FTT and the duality theorem 
is purely algebraic \cite{Catani:2008xa}. We explicitly illustrate 
it by considering 
the massless two-point function ${L}^{(2)}(p_1, p_2)$.
Its dual representation is
\beeq
\label{twodual1}
&& \!\!\!\!\!\!\!\!\!\!\!
{\widetilde L}^{(2)}(p_1, p_2) = I^{(1)}(p_1) + I^{(1)}(p_2) \nn \\
&& \!\!\!\!\!\!\!\!\!\!\!
= \int_q \, \td(q) \left( 
\frac{1}{2qp_1+p_1^2-i0 \,\eta p_1} + (p_1 \leftrightarrow p_2) \right)~.
\nn \\ \eeeq
Inserting Eq.~(\ref{dovsfp}) in Eq.~(\ref{twodual1}), we 
obtain 
\beeq
\label{twodual2}
&& \!\!\!\!\!\!\!\!\!\!\!
{\widetilde L}^{(2)}(p_1, p_2) = L^{(2)}_{1-{\rm cut}}(p_1, p_2) \nn \\
&& + \left[ \theta(\eta p_1) + \theta(\eta p_2)\right]
\, L^{(2)}_{2-{\rm cut}}(p_1, p_2)~. 
\eeeq 
Owing to momentum conservation (namely, $p_1+p_2=0$) $\theta(\eta p_1) + \theta(\eta p_2)=1$, 
and then the dual and the FTT  representations of the two-point 
function are equivalent. 

The proof of the equivalence in the case of higher $N$-point functions proceeds in a similar 
way \cite{Catani:2008xa},  the key ingredient simply being the constraint of
{\em momentum conservation}.

\section{MASSIVE PARTICLES, UNSTABLE PARTICLES}

We have so far considered massless propagators. The extension to include 
propagators with finite mass $M_i$ is straightforward, as long as $M_i$ 
is real. The massless on-shell delta function
$\td(q_i)$ is replaced by
\beq
\label{dmass}
\td(q_i;M_i) = 2 \pi \, i \;\delta_+(q_i^2-M_i^2)~,
\eeq
when the $i^{\rm th}$ line of the loop is cut to obtain 
the dual representation. 
The $i0$ prescription of the dual propagators 
is not affected by real masses. The corresponding dual 
propagator is
\beq
\frac{1}{q_j^2 - M_j^2 - i0 \,\eta (q_j-q_i)}~.
\eeq

In field theories with unstable particles,
a Dyson summation of self-energy insertions is required to properly treat the 
propagator $G_C$ of those particles.  
This produces finite-width effects, introducing finite 
imaginary contributions in those propagators.
A typical form of the propagator $G_C$ 
(such as in the complex-mass scheme \cite{cms}) is 
\beq
\label{comp}
G_C(q;s) = \frac{1}{q^2-s}~,
\eeq
where $s$ denotes the complex mass of the unstable 
particle $(s = {\rm Re \;} s + i \, {\rm Im \;} s$,
${\rm Re \;} s > 0 > {\rm Im \;} s)$.

The complex-mass propagators produce poles that are located far 
off the real axis. Thus, when using the Cauchy theorem, as in Eq.~(\ref{ln4}), 
the duality relation is built up from two contributions
\beq
\label{repdt}
L^{(N)} = - \left( {\widetilde L}^{(N)} + {\widetilde L}_C^{(N)} \right)~.
\eeq
Here, ${\widetilde L}^{(N)}$ denotes the terms that correspond to the residues 
at the poles of the Feynman propagators of the loop integral, while 
${\widetilde L}_C^{(N)}$ denotes those from the poles of the complex-mass 
propagators. 

In other schemes, the propagator of an unstable particle can have 
a form that differs from Eq.~(\ref{comp}). 
One can introduce, for instance, a complex mass
that depends on the momentum $q$ of the propagator, i.e. $s(q^2)$, or even 
a non-resonant component in addition to the resonant contribution. 
Independently of its specific form, the propagator $G_C$ of the unstable 
particle produces singularities that are located at a {\em finite} imaginary 
distance from the real axis in the $q_0$ complex plane. 
Owing to this finite displacement, the structure of the duality relation
(\ref{repdt}) is valid in general, although the explicit form of 
${\widetilde L}_C^{(N)}$ depends on the actual expression of 
the propagator $G_C$.

\section{GAUGE POLES}

The quantization of gauge theories requires the introduction of a gauge-fixing
procedure, which specifies the spin polarization vectors of the 
gauge bosons and the content
of possible compensating fictitious 
particles (e.g.~the Faddeev--Popov ghosts in unbroken 
non-Abelian gauge theories,
or the would-be Goldstone bosons in spontaneously broken gauge theories).
The Feynman propagators of the fictitious particles are treated exactly in the
same way as those of physical particles when deriving (applying)
the duality relation. 

The propagators of the gauge particles, however, 
can introduce extra unphysical poles.
The general form of the polarization tensor of a spin-one gauge boson is
\beq
\label{polten}
d^{\mu \nu}(q) = - g^{\mu \nu} + (\zeta -1) \; {\ell}^{\mu \nu}(q) \,G_G(q) 
\;\;.
\eeq
The second term on the right-hand side is absent only in the 
't~Hooft--Feynman gauge $(\zeta=1)$. In any other gauge, 
the tensor ${\ell}^{\mu \nu}(q)$ propagates 
longitudinal polarizations. Its specific form is not relevant in the context of
the duality relation. Indeed, ${\ell}^{\mu \nu}(q)$
has a polynomial dependence on the momentum $q$ 
and, therefore, it does not interfere   
with the residue theorem. The factor $G_G(q)$ 
(`gauge-mode' propagator), however, has a potentially dangerous
non-polynomial dependence on $q$, and it
can introduce extra (i.e. in addition to the poles of the associated Feynman
propagator)
poles with respect to the momentum variable $q$.  
A typical example of `gauge poles' are those located at $q \cdot n=0$ in 
the case of axial gauges (here $n_\mu$ is the axial-gauge vector).

The presence of gauge poles in $G_G(q)$ can modify the form 
of the duality relation. In general, one can expect that, 'cutting' the loop
(i.e. applying the residue theorem to the loop integral), one has to explicitly
include the absorptive contribution from the gauge-mode propagators 
in addition to the customary single-cut contribution from the Feynman
propagators. Moreover, this additional contribution would have  
a different form in different gauges.
 
The impact of gauge poles on the duality relation is discussed in detail
in \cite{Catani:2008xa}.
The duality relation in the simple form presented here
(i.e. with the inclusion of the sole single-cut terms from the Feynman
propagators) turns out to be valid~\footnote{Of course, the duality relation 
is obviously valid in the 't~Hooft--Feynman gauge, where there are no gauge 
poles.} in spontaneously-broken gauge theories in the unitary gauge,
and in unbroken gauge theories in physical gauges specified by a gauge vector 
$n^\nu$, {\em provided} the dual vector $\eta^\mu$ is chosen 
such that  $n\cdot \eta=0$. This excludes gauges where $n^\nu$ 
is time-like. Note that the validity of the duality relation in this form
does not imply that the loop integral receives no extra contributions from
the gauge poles. It simply implies that these contributions
arise after the phase-space integration of the corresponding single-cut 
integrals. 

\vspace*{-2mm} 
\section{DUALITY AT THE AMPLITUDE LEVEL}
\vspace*{-2mm}
The duality relation can be applied to evaluate not only basic one-loop integrals 
$L^{(N)}$ but also complete one-loop quantities ${\cal A}^{({\rm 1-loop})}$
(such as Green's functions and
scattering amplitudes). 
The analogue of Eqs.~(\ref{ldt}) and (\ref{dcut})
is the following duality relation \cite{Catani:2008xa}:
\beq
\label{adt}
{\cal A}^{({\rm 1-loop})} = - \;
{\cal \widetilde A}^{({\rm 1-loop})}
\;\;.
\eeq
The expression ${\cal \widetilde A}^{({\rm 1-loop})}$ on the right-hand side
is obtained from ${\cal A}^{({\rm 1-loop})}$
in the same way as ${\widetilde L}^{(N)}$ is obtained from ${L}^{(N)}$:
starting from any Feynman 
diagram in ${\cal A}^{({\rm 1-loop})}$, we consider all possible
replacements of each Feynman propagator $G(q_i)$ in the loop with
the cut propagator $\td(q_i;M_i)$, and then we replace
the uncut Feynman propagators with dual propagators. All the other factors
in the Feynman diagrams are left unchanged in going from 
${\cal A}^{({\rm 1-loop})}$ to ${\cal \widetilde A}^{({\rm 1-loop})}$.

Equation (\ref{adt}) establishes a correspondence between the one-loop Feynman
diagrams contributing to ${\cal A}^{({\rm 1-loop})}$ and the
tree-level Feynman diagrams contributing to 
the phase-space integral in ${\cal \widetilde A}^{({\rm 1-loop})}$.
How are these tree-level Feynman diagrams related
to those contributing to the tree-level expression~\footnote{
Here ${\cal A}^{({\rm tree})}$ exactly denotes the tree-level counterpart of 
${\cal A}^{({\rm 1-loop})}$.} ${\cal A}^{({\rm tree})}$?
The answer to this question is mainly a matter of combinatorics of 
Feynman diagrams. If ${\cal A}^{({\rm 1-loop})}$ is an 
off-shell Green's function, the 
phase-space integrand in ${\cal \widetilde A}^{({\rm 1-loop})}$ is directly
related to ${\cal A}^{({\rm tree})}$ \cite{Catani:2008xa}. 
In a sketchy form,  we can write:
\vspace{-1mm}
\beeq
\label{adgen}
&& \!\!\!\!\!\!\!\!\!\!\!
{\cal A}_N^{({\rm 1-loop})}(\dots) \nn \\ && 
\sim \int_q \, \sum_{P} \, \tilde\delta_+(q;M_P) \, 
{\cal \widetilde A}_{N+2}^{({\rm tree})}(q,-q,\dots)~,
\eeeq 
where $\sum_{P}$ denotes the sum over the types of particles and antiparticles 
that can propagate in the loop internal lines, and 
${\cal \widetilde A}^{({\rm tree})}$ simply differs from 
${\cal A}^{({\rm tree})}$ by the replacement of dual and Feynman propagators.
If the tree-level Green's function ${\cal A}_{N+2}^{({\rm tree})}$  
with  $N+2$ external particles 
is known, it can be reused in Eq.~(\ref{adgen}) to calculate 
the corresponding one-loop Green's function with $N$ external particles.

The extension of Eq.~(\ref{adgen}) to scattering amplitudes 
requires a careful treatment of 
the on-shell limit of the corresponding Green's functions 
\cite{Catani:2008xa}.

\section{SUMMARY}

We have illustrated a duality relation between loops and trees.
One-loop integrals are written 
in terms of single-cut phase-space integrals, with 
propagators regularized by a new Lorentz-covariant 
$i0$ prescription.
This simple modification of the Feynman propagators
compensates for the absence of multiple-cut contributions that 
appear in the FTT. The duality relation has been extended 
from Feynman integrals to off-shell Green's functions.
Work is in progress \cite{meth,tanju,inprep} on applications 
to the computation of
one-loop scattering amplitudes and NLO cross sections.

\end{document}